# Coherent Photonic Crossbar as a Universal Linear Operator

George Giamougiannis,[1,*] Apostolos Tsakyridis,[1] Yangjin Ma,[2] Angelina Totovic,[1] David Lazovsky,[2] Nikos Pleros[1]

[1]*Department of Informatics, Center for Interdisciplinary Research & Innovation, Aristotle University of Thessaloniki, Thessaloniki, Greece*
[2]*Celestial AI, 100 Mathilda Place, Suite 170, Campbell, CA 95008, United States*
**Corresponding author: giamouge@csd.auth.gr*

**Linear optics aim at realizing any real- and/or complex-valued matrix operator via optical elements, addressing a broad field of applications in the areas of quantum photonics, microwave photonics and optical neural networks. The transfer of linear operators into photonic experimental layouts typically relies on Singular Value Decomposition (SVD) techniques combining meshes of cascaded 2×2 Mach-Zehnder Interferometers (MZIs), with the main challenges being the precision in the experimental representation of the targeted matrix, referred to as fidelity, and the overall insertion loss. We demonstrate a novel interferometric coherent photonic crossbar architecture (Xbar) that demarcates from state-of-the-art SVD implementations and can realize any linear operator, supporting full restoration of the loss-induced fidelity. Its novel interferometric design allows for the direct mapping of each matrix element to a single, designated Xbar node, bringing down the number of programming steps to only one. We present the theoretical foundations of the Xbar, proving that its insertion losses scale linearly with the node losses as opposed to the exponential scaling witnessed by the SVD counterparts. This leads to a matrix design with significantly lower overall insertion losses compared to SVD-based schemes when utilizing state-of-the-art silicon photonic fabrication metrics, allowing for alternative node technologies with lower energy consumption and higher operational speed credentials to be employed. Finally, we validate that our Xbar architecture is the first linear operator that supports fidelity restoration, outperforming SVD schemes in loss- and phase-error fidelity performance and forming a significantly more robust layout to loss and phase deviations.**

## 1. INTRODUCTION

There has been an increasing interest in linear optics during the last years, aiming at deployment of programmable universal multiport interferometers capable of performing any linear transformation on any given set of input modes [1-3]. In this effort, the ultimate goal is the experimental realization of any real- or complex-valued matrix operator solely via optical elements, allowing for the exploitation of all speed-, size- and energy-related advantages of optics in the rapidly advancing fields of neuromorphic photonics [4-7], microwave photonics [8-9], machine learning [3,10] and quantum photonics [2,11] among other. Optical linear transformations have been so far the stronghold of matrix designs that rely, mainly, on the Singular Value Decomposition (SVD) technique, as has been first proposed by Miller [1]. This scheme extends along the factorization of any matrix $D$ in the form of $D=U\Sigma V^{\dagger}$, where $U$ and $V$ are unitary matrices implemented in the optical domain and $\Sigma$ is a diagonal matrix that can be optically realized via a column of variable attenuators interleaved between the two unitary matrix layouts. As such, the problem of enabling any matrix representation in the optical domain via the SVD approach turns into the constituent problem of optically realizing any unitary matrix. The field of optical unitary matrix operators was pioneered in the seminal work of Reck et. al. [12], where a unitary matrix decomposition scheme based on elementary 2×2 unitary elements was adopted. With the 2×2 unitary element inherently offered by a single lossless 2×2 Mach-Zehnder Interferometer (MZI) in the optical domain, along with two Phase Shifters (PSs), Reck et. al. proved that any N×N unitary operator can be implemented via a triangular mesh of N(N-1)/2 MZI nodes. This was, recently, validated also experimentally in a silicon-integrated Programmable Nanophotonic Processor (PNP) that was successfully employed in an all-optical neural network for vowel recognition [3] and a silicon optical neural chip for the implementation of complex valued neural networks [7], highlighting the advantages of linear optics in the rapidly emerging field of neuromorphic photonics. Moving towards an improved unitary matrix scheme, Clements et. al. [13] recently presented a more compact rectangular mesh architecture that requires the same number of 2×2 MZI nodes but yields half the optical depth compared to the Reck design, supporting in this way a more loss-balanced and loss-error-tolerant design.

Both Reck and Clements architectures have been based on the assumption of lossless 2×2 MZIs, as their elementary constituent nodes, towards realizing unitary matrix operators. Yet, the reality is that non-ideal optical elements will be utilized when these architectures will be transferred into experimental layouts, implying that lossy elementary MZI nodes and eventually also imperfect phase conditions will definitely affect the device and, consequently, the system performance. Given that both Reck and Clements designs rely on cascaded stages of 2×2 MZI nodes, their overall insertion loss will scale exponentially with the MZI losses, eventually limiting the circuit size due to the loss build-up. Moreover, the employment of lossy optical MZI nodes yields a loss-imbalanced configuration that translates directly into reduced fidelity metrics. Although the Clements layout is inherently more loss-resistive than the Reck design, both schemes can neither sustain a 100% fidelity nor practically support any fidelity restoration mechanism. This indicates that both designs come with the demand of ultra-low-loss MZI node technology, in order to retain reasonable insertion losses and high fidelity performance, forming one of the main reasons for the use of low-loss thermo-optic MZI elements in the experimental deployments reported so far [3,7].

These effects become even more pronounced when any real- and/or complex-valued linear transformation via the SVD factorization is targeted, where two concatenated unitary matrix layouts are required. Fidelity degradation and high insertion losses will form an important drawback especially when higher matrix dimensions are targeted, restricting even more the employment of alternative MZI PS technologies that may eventually equip the matrix operator with additional energy, functionality or speed advantages. For example, both Clements and Reck designs can hardly cope with the losses of emerging, yet powerful, photonic technologies, like Phase-Change Materials (PCMs) [14-15] and electro-optic phase shifting structures [16], which typically have insertion loss values beyond 1dB per element but offer significant benefits in terms of energy consumption and speed, compared to their thermo-optic PS counterparts.

In this paper, we propose a novel interferometric coherent photonic crossbar architecture (Xbar) that can realize any real- and/or complex-valued matrix operator and outperforms SVD factorization-based schemes in several key aspects. First, it allows for natural one-to-one mapping of each matrix element into the designated Xbar node, avoiding cascaded nodes for a single matrix element representation and yielding a linear dependence of the total insertion losses on the individual node losses. This can significantly reduce insertion losses compared to SVD-based layouts when exploiting state-of-the-art silicon photonic technology metrics, with the loss benefits becoming even more pronounced when a higher loss node technology or higher matrix dimensions are targeted. Second, it follows a modular design that can be tailored either for ensuring a loss-balanced performance among all columns or for associating every matrix column with a constant loss factor, leading in this way to a completely restorable fidelity performance. This outlines a highly loss-tolerant design that can always precisely implement the targeted linear transformation, yet being flexible in terms of MZI node technology, as it can use even higher-loss PS technologies for enriching the matrix deployment with energy, speed and/or functionality benefits. Third, its non-cascaded, one-to-one mapping architecture makes the design significantly more robust to fabrication errors caused by phase mismatches, restricting the error only to a single matrix element, as opposed to error being accumulated to the multiple matrix elements due to non-bijective correspondence between matrix element and photonic node element in cascaded architectures. Last but not least, it can be programmed in just a single-step compared to the N(N-1)/2, taking advantage of its modular design that exploits decoupled Xbar nodes for matrix element representation. The proposed Xbar architecture effectively extends our previous work on coherent dual-IQ modulator-based vector dot-product engine [17] towards a full vector-by-matrix multiplication interferometric layout, employing multiple parallel columns within a split-and-recombine configuration.

## 2. BACKGROUND

The prevailing method for constructing any real and/or complex-valued matrix operator in the optical domain is based on the SVD factorization technique. This factorization relies on the decomposition of the matrix $D$ into a product of the form UΣV†, where $\Sigma$ is a diagonal matrix and $U$ and $V$ are two unitary matrices, with $V^\dagger$ being the conjugate transpose of $V$. To this end, the optical deployment of the SVD-based scheme can take advantage of the unitary matrix operator implementations that have been so far proposed in the optical domain [12-13] in order to construct the $U$ and $V^\dagger$ matrices, incorporating also an additional column of optical attenuator elements in between the two photonic unitary matrix designs for realizing the $\Sigma$ diagonal matrix. Given that the Clements unitary matrix design has been proven, so far, to be the optimal layout in terms of optical depth and loss-induced fidelity [13], we consider as the optimal SVD-based photonic linear operator the layout that adopts the Clements unitary matrix architecture for both its $U$ and $V^\dagger$ operators and we will refer to it as the SVD-Clements architecture. This implies that an N×N optical SVD-Clements design will require N(N-1)/2 programming steps for its configuration [18], since the two constituent N×N Clements unitary matrices can be programmed in parallel within N(N-1)/2 steps.

Fig. 1(a) illustrates an example of an N×N SVD-Clements architecture for N=6, when operating on an N-element input vector $X$. The values of the input vector, $x_r$, with $r = 1,2,\dots,N$ denoting the row index, are imprinted onto the optical beams originating from the incoming optical continuous wave (CW) signal $E_{in}$, split equally in terms of power by a 1:N optical splitter. Vector by matrix multiplication starts by modulated optical signals, $x_r E_{in}/N$, entering the N×N unitary operator $V^\dagger$ that relies on the Clements rectangular mesh of 2×2 MZI-based nodes, with every 2×2 MZI node comprising two 3dB couplers and two PSs $\theta$ and $\varphi$, as depicted in the yellow inset of Fig. 1(a). After exiting the optical layout of the $V^\dagger$ unitary operator, the $N$ parallel optical beams enter the optical diagonal matrix $\Sigma$ actualized by a single column of $N$ variable optical attenuators (VOAs). Every attenuator can be easily realized via an MZI element where only one of its inputs connects to the preceding $V^{\dagger}$ matrix and only one of its two outputs connects to the succeeding $U$ matrix, as shown in the dashed-line inset of Fig. 1(a). The N×N $U$ matrix design, based again on the Clements architecture, employ the layout and basic building blocks that are used also for the $V^{\dagger}$ matrix operator. In this way, the $N$ optical beams emerging at the output waveguides of the $U$ matrix design yield an $N$-element optical output vector $Y$ with $Y = U\Sigma V^{\dagger}X$.

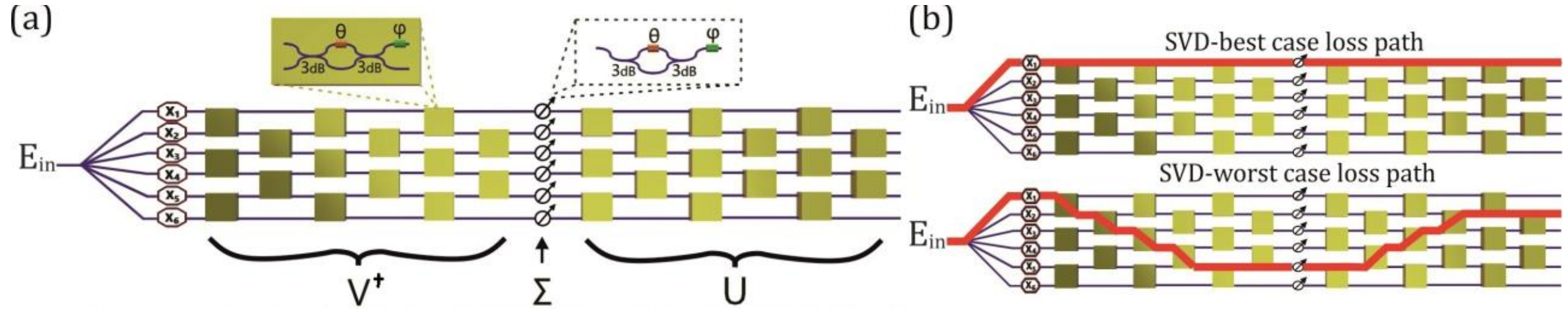

**Fig. 1.** (a) The SVD-Clements architecture with the Clements unitary rectangular mesh utilized for the $V^{\dagger}$ and $U$ matrix deployment within the SVD implementation. A single node of Clements implementation and the design of the attenuators comprising the $\Sigma$ matrix are depicted via insets. (b) Indicative examples of best- (top) and worst-case (bottom) loss signal paths in the SVD-Clements design.

Each of the N×N $U$ and $V^{\dagger}$ unitary operators requires a total number of N(N-1)/2 MZI nodes and has a maximum optical depth [13] of $N$, with its shortest path equalling $\lfloor N/2 \rfloor$, where $\lfloor x \rfloor$ denotes the lower integer bound of $x$. This implies that the entire N×N SVD-Clements layout will employ a total number of $N^2$ MZI nodes, out of which, N(N-1) come from 2×2 MZI nodes of both $U$ and $V^{\dagger}$ matrices added to the $N$ MZI nodes used for the $\Sigma$ diagonal matrix. At the same time, the depth of the shortest optical path of the SVD-Clements design will be $(2\lfloor N/2 \rfloor + 1)$ and will be named hereinafter as SVD-best-case loss path, while the maximum optical depth will equal $(2N+1)$ and will be hereinafter referred to as SVD-worst-case loss path, with both the best- and the worst-case loss paths illustrated schematically in Fig. 1(b).

When 2×2 MZI nodes are considered to be lossless, both $U$ and $V^{\dagger}$ matrix implementations yield indeed unitary transformations. Assuming that the input modulators $x_r$ are also lossless and operate at their transparency region, then the overall insertion losses of the SVD-Clements architecture are solely dictated by the attenuation levels enforced at its $\Sigma$ diagonal matrix, which, in turn, depend on the specific singular values of the targeted matrix. Decoupling the insertion losses of the overall SVD-Clements architecture from the matrix-specific attenuation values required in the $\Sigma$ column, the MZI nodes that form the $\Sigma$ diagonal matrix are also considered to perform in their transparency region, i.e., with no loss penalty. In this case, the entire SVD-Clements design can be indeed considered as lossless and the output power obtained at each individual output waveguide is exclusively defined by the targeted matrix dimensions through the 1/N splitting ratio.

The outcome becomes different when the constituent MZI nodes are not treated as lossless, as will, in fact, be the case in a practical implementation. Let us define as $l_{coup} \leq 1$ and $k \leq 1$ the electric field transmission coefficients of a 3dB coupler and a PS, respectively, where $IL_{coup,dB} = -10\log_{10}(l_{coup}^2)$ and $IL_{PS,dB} = -10\log_{10}(\mathrm{k}^2)$ denote the corresponding optical power insertion losses. In each MZI-node, where two 3dB couplers and two PSs $\theta, \varphi$ are employed, the total electric field transmittivity factor is $T_{node} = l_{coup}^2 k^2 \leq 1$ and the total MZI-node insertion loss $IL_{node,dB} = -10\log_{10}(T_{node}^2)$. Assuming transparent and lossless modulators $x_r$, we can define the overall electric field transmission coefficient between the matrix output and input as $f$, with $f^2$ denoting the power transmission coefficient. Taking into account that the electric field emerging at a single matrix output waveguide will be the coherent sum of multiple electrical fields that will have propagated through different paths within the matrix, $f^2$ will be bounded between the transmittivity of the longest and shortest optical paths within the SVD-Clements architecture respectively, $f^2 \in \left[\frac{1}{N}\left(T_{node}^2\right)^{(2N+1)}, \frac{1}{N}\left(T_{node}^2\right)^{\left(2\left\lfloor\frac{N}{2}\right\rfloor+1\right)}\right] \leq \frac{1}{N}$. The upper bound of $f^2$ corresponds to the optical power transmittivity in the SVD-best-case path loss scenario, where the signal travels through $(2\lfloor N/2 \rfloor + 1)$ nodes, as depicted in Fig. 1(b)(top), while its lower bound stands for the optical power transmittivity in the SVD-worst-case path loss scenario, where the signal propagates through the maximum number of nodes $(2N+1)$, as illustrated in Fig. 1(b)(bottom). This indicates that the overall insertion losses of the SVD-Clements architecture depend exponentially on the MZI-node insertion loss $T_{node}^2$ regardless of the path taken, or, in terms of dB, linearly on $IL_{node,dB}$, with only the slope being defined by the taken path. Expressing the total insertion losses in dB for the two extreme cases yields:

$$IL_{SVD-Clements,bc,dB} = 10\log_{10}(N) + \left(2\left\lfloor\frac{N}{2}\right\rfloor + 1\right) IL_{node,dB} \tag{1}$$

for the best-case scenario and

$$IL_{SVD-Clements,wc,dB} = 10\log_{10}(N) + (2N+1) IL_{node,dB} \tag{2}$$

for the worst-case scenario.

Besides this exponential scaling, the employment of MZI nodes with non-zero insertion losses is also expected to degrade the fidelity of the experimentally realized matrix layout, being the result of non-balanced losses through different paths. Each of the constituent unitary matrices relies on the Clements design that has been shown to yield a reduced fidelity performance when MZI nodes of nonzero insertion losses are used, with fidelity degradation becoming stronger with increasing MZI losses and circuit size [13]. To this end, the employment of the Clements design in the respective SVD-Clements architecture for realizing any linear and not only unitary transformations, as is the ultimate target in the field of linear optics, is expected to yield even stronger loss-induced fidelity degradation since two cascaded unitary layouts along with an additional diagonal matrix $\Sigma$ are required.

## 3. COHERENT CROSSBAR AS A LINEAR OPERATOR

Photonic crossbars have been so far well-known from networking and switching sectors, with their main drawback being their differential path losses, enforcing the use of non-coherent multiwavelength input signals with unequal power levels when their utilization as a matrix operator is targeted [19-20]. Here, we propose a coherent Xbar layout, able to operate using a single wavelength, allowing for wavelength multiplexing to become an additional degree of freedom to be used for other purposes, e.g., throughput boosting through operation parallelization. Our Xbar architecture is an extension and a generalization of our recently reported coherent vector dot-product engine, operating as a coherent linear neuron, relying on dual-IQ modulation cell as its main building block [17].

An N×M Xbar layout when multiplying an N-element input vector with $N = 2^n$ and $n, M \in \mathbb{N}$ is depicted in Fig. 2. It comprises a 1:N front-end splitter that is followed by the input vector $X$ modulation stage and the main block of the Xbar. The latter is formed by $M$ columns, with every column comprising an array of the Xbar nodes followed by a back-end combination-and-forwarding stage (BCFS). An input CW signal $E_{in}$ is injected into the front-end splitter and is then equally split, in terms of power, in $N$ identical constituent beams via $log_2(N)$ cascaded 3dB Y-junction splitting couplers. These $N$ CW signals are then modulated by their corresponding input amplitude modulator in order to form the optical data signal vector $X = [x_1, x_2, \dots, x_N]^T$, with its $r$-th element represented by $x_r$, which then enters the Xbar main block via its $r$-th row. Each of these signals is then split via a $\xi_1^2 : t_1^2$ optical splitter, so that the part of the $r$-th row signal that gets coupled to the $\xi_1^2$ splitter output enters the Xbar node (r,1). The remaining part of the $r$-th row signal, that gets coupled to the $t_1^2$ splitter output, continues its horizontal route along the row until reaching the next Xbar column, where an additional $\xi_2^2 : t_2^2$ optical splitter will again split the signal into a portion entering the 2nd column of the Xbar (proportional to $\xi_2^2$) and a portion that gets forwarded to its 3rd column (proportional to $t_2^2$). This process is repeated at every Xbar node (r,c) via a respective $\xi_c^2 : t_c^2$ optical splitter until reaching the last column $M$, which includes no splitters, but rather relies on the (M-1)-th column splitter $\xi_{M-1}^2 : t_{M-1}^2$ to receive the optical signal from its $t_{M-1}^2$output. In other words, column $M$ could be described by a virtual splitter with $\xi_M^2 = 1$ and $t_M^2 = 0$.

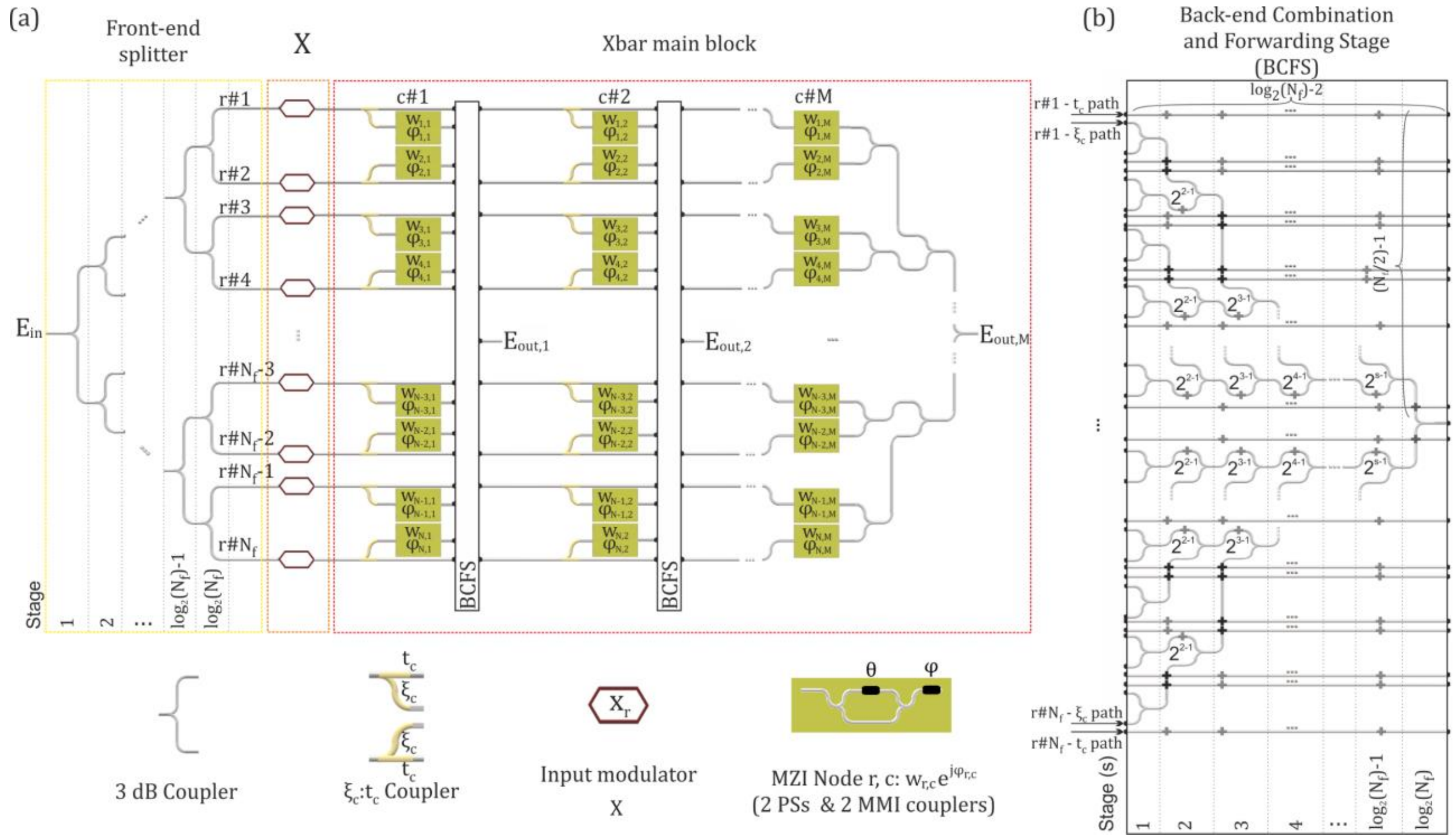

**Fig. 2.** a) The N×M Xbar core as a vector-by-matrix multiplication engine that comprises a front-end splitter, the $X$ input vector modulation stage and the NxM Xbar main block, which in turn includes a back-end combination and forwarding stage (BCFS) at every column; b) the detailed layout of the BCFS stage.

Each Xbar node at the $r$-th row and $c$-th column of the Xbar comprises an amplitude modulator, with its transfer function being proportional to $w_{r,c}$ and a PS $\varphi_{r,c}$, so that the $r$-th row signal part coupled through the $\xi_c^2$ branch and travelling vertically along the column $c$ gets multiplied by a factor $w_{r,c}e^{j\varphi_{r,c}}$. Taking into account that $w_{r,c}$ is non-negative, this factor is a complex value with a norm of $w_{r,c}$ and a phase $\varphi_{r,c}$, implying that every Xbar node (r,c) can represent any real or complex number at its (r,c)-th matrix element. The experimental realization of an Xbar node can be accomplished through the employment of a tunable MZI that has only one of its 2 inputs and one of its 2 outputs connected to the circuit and acts as a VOA for defining the $w_{r,c}$ value, followed by a tunable PS for defining the value $\varphi_{r,c}$, as shown in the legend of Fig. 2(a).

After exiting the Xbar node, the signal enters the BCFS of its corresponding column $c$ that is responsible for performing the following two functions: a) forward the signal part emerging at the $t_c^2$ coupling port of the $\xi_c^2 : t_c^2$ optical splitter to the next column via a passive optical waveguide section, b) allow the signals that exit the intra-column nodes from all Xbar rows to coherently recombine and produce the column output signal $E_{out,c}$. The BCFS is shown in detail in Fig. 2(b), with the coherent recombination being accomplished via a mirror version of the 1:N front-end splitter implemented through a cascade of 3dB Y-junction combiner stages. In this way, the $N$ signals can recombine sequentially in clusters of two at every combination stage until reaching the single waveguide output. Waveguide crossings are required at every column $c \in [1, \mathrm{M}-1]$ between successive combination stages until reaching the single column read-out in order to overcome the row waveguides that lead to the next columns acting as the forwarding section. It should be recognized that a different number of crossings is required at every row of each column, with the maximum being $log_2(N) - 2$ at rows $(N/2) - 1$ and $(N/2) + 2$, for $N > 2^2$, which typically is the case. If $N = 2^2$, only a single crossing appears in rows $N/2 = 2$ and $N/2 + 1 = 3$, whereas for $N = 2$ no crossings are required. In summary, for $N \geq 2^2$, we have a maximum number of waveguide crossings that amounts to $\max\{1, \log_2(N) - 2\}$. In order to ensure a loss-balanced forwarding section, every row waveguide in the BFCS is equipped with dummy crossings up to a total number of $\max\{1, \log_2(N) - 2\}$ in each column $c$, for $N \geq 2^2$, as shown in Fig. 2(b). At the same time, the need to ensure balanced losses between the recombining signals within every column leads to the use of $2^{s-1}$ dummy waveguide crossings at the respective input port of every Y-junction combiner in the $s$-th combination stage, with $s = 2,3, \dots, log_2(N) - 1$. Together with the final single crossing at the last $log_2(N)$ stage, the total number of crossings along the recombination path that each row-signal encounters amounts to $1 + \sum_{s=2}^{\log_2(N)-1} 2^{s-1} = N/2 - 1$, for $N \geq 2^2$. In the exception of $N = 2$, no crossings will exist, similar as along the forwarding waveguides.

Up to this point, the Xbar architecture was described with power-of-2 number of rows $N$, but it should be stressed that it can also be implemented in any N×M matrix dimension, even in the case of $N$ values that are not a power of 2. This can be realized by simply deactivating or ignoring the number of rows that correspond to the difference between $N$ and $N_f$ within a $N_f$×M Xbar design, where $N_f = 2^n$ and $n = \lceil log_2(N) \rceil$, with the brackets denoting the lower integer bound. For example, for an intended N×M layout where N=8 and M=10, then $N_f = N = 8$. In the case of N=5, then $N_f$ will remain equal to 8 according to the relationship $N_f = 2^{\lfloor log_2(5) \rfloor + 1} = 8 > N$, but three of its rows will remain inactive. This indicates that part of the input signal power will be wasted resulting to additional insertion losses; however, this loss penalty will never exceed 6 dB given that $N$ will be always between $2^{n-1}$ and $2^n$. Finally, it is important to note that, since every Xbar node represents an individual matrix element, all Xbar nodes can be configured independently and in parallel to each other requiring just a single-step programming operation for the entire matrix, offering a significant configuration time benefit compared to the respective N(N-1)/2 steps needed to program an SVD-Clements layout.

Focusing on a single, $c$-th column of the Xbar, it can be clearly seen how its operation represents a generalization of our coherent vector dot-product engine described in Ref. 17. The modulated input signal vector $X$ is fed into every of the Xbar's columns exploiting the $\xi_c^2 : t_c^2$ optical splitter stage, element-wise multiplied by the $c$-th column vector of the Xbar matrix, and coherently summed at the BCFS stage to yield the result of dot-product multiplication, $E_{out,c}$, at the output. Every individual Xbar column connects to the incoming optical signal vector $X$ as described above, effectively implementing generalized version of the layout described in Ref. 17. Considering again that the insertion loss of each MZI-node equals $IL_{node,dB} = -10\log(T_{node}^2)$, as in the case of SVD-Clements design, the electric field at every column output of the Xbar can be obtained through:

$$E_{out,c} = \alpha T_{node} L_c \frac{1}{N_f} \left( \prod_{q=1}^{c-1} t_q \right) \xi_c \left( \sum_{r=1}^{N} x_r w_{r,c} e^{j\varphi_{r,c}} \right) E_{in} \tag{3}$$

where $\alpha \leq 1$ is the electric field transmission coefficient of the input amplitude modulator $x_r$, when operated in transparency, while $N_f$ and $N$ are the row-dimensions of the Xbar layout and the targeted matrix, respectively. In case of the last column, no splitting stage exists and signal is only passed through the column itself, implying $\xi_M = 1$. The coefficient $L_c \leq 1$ is defined as the total passive circuitry transmission coefficient, denoting the transmission through all front-end and back-end 3dB couplers (Y-splitters and Y-combiners, respectively), all $\xi_c^2 : t_c^2$ couplers and all waveguide crossings within the signal's path from Xbar input to $c$-th column output. For $N > 2$, it equals:

$$L_c = \begin{cases} l_{coup}^{2\log_2(N_f)}\, l_\xi^c\, l_x^{\frac{N_f}{2}-1+(c-1)\max\{1,\log_2(N_f)-2\}}, & c \in [1, M-1] \\ l_{coup}^{2\log_2(N_f)}\, l_\xi^{M-1}\, l_x^{(M-1)\max\{1,\log_2(N_f)-2\}}, & c = M \end{cases} \tag{4}$$

where $l_\xi \leq 1$ and $l_x \leq 1$ are the electric field transmission coefficients of a single $\xi^2{:}\,t^2$ directional coupler and a waveguide crossing, respectively. If $N = 2$, no waveguide crossings will exist, reducing the expression for $L_c$ to the case of $l_x = 1$, and raising the coefficient $L_c$ itself closer to the ideal value of 1. From this point onward, we focus on the case of $N > 2$, a condition under which we anticipate Xbar to typically operate. A detailed derivation of Eq. (3), together with the accompanying explanations, is given in Section A of the Supplementary material.

Equation (3) verifies that the output of each column is proportional to the linear weighted sum of the inputs $x_r$ that is embraced in the parentheses, successfully yielding the vector dot-product between the Xbar column entries and the input column vector $X$. Based on the column outputs given by Eq. (3), we define the Xbar associated N×M transfer matrix, i.e., a weight matrix, normalized to lossless case as:

$$W = \begin{bmatrix} w_{1,1}e^{j\varphi_{1,1}} & \cdots & w_{1,M}e^{j\varphi_{1,M}} \\ \vdots & \ddots & \vdots \\ w_{N,1}e^{j\varphi_{N,1}} & \cdots & w_{N,M}e^{j\varphi_{N,M}} \end{bmatrix} \tag{5}$$

with each Xbar node being described by $w_{r,c}e^{j\varphi_{r,c}}$ where $r = 1,2,\dots,N$ and $c = 1,2,\dots,M$. The output electrical field vector $E_{out} = \left[E_{out,1}, E_{out,2}, \dots, E_{out,M}\right]^T$ can now be expressed in a matrix-equation form as:

$$E_{out} = (WP)^T X E_{in} = P^T W^T X E_{in} \tag{6}$$

with $P$ being the diagonal M×M transmission matrix encompassing all the transmission factors corresponding to the insertion losses and splitting ratios $\mathrm{P} = \mathrm{diag}(p_1, p_2, \dots, p_M)$, where the real elements along its diagonal equal:

$$p_c = \alpha T_{node} L_c \frac{1}{N_f}\left(\prod_{q=1}^{c-1} t_q\right)\xi_c \tag{7}$$

More details are given in section A of Supplementary material.

## 4. INSERTION LOSS ANALYSIS

The overall insertion loss at the $c$-th output column of the Xbar can be calculated using the respective transmission factor $p_c$. In order to do so, we assume transparent operation of all input modulators, $x_r = 1$, as well as all Xbar nodes with $w_{r,c}e^{\varphi_{r,c}} = 1$, so as to exclude the matrix-value-enforced node losses, as was done in the case of the SVD-Clements architecture. A simple insertion loss analysis can be carried out when considering a loss-balanced N×M matrix, by which we assume that all of Xbar's column outputs have the same insertion loss. Following the analysis described in detail in Section B of the Supplementary material, we show that the $\xi_c^2$ coupling coefficients can be selected in such a manner to yield identical transmission factors at every column, i.e., $p_c = p_1$ for $c \in [2, \mathrm{M}]$, effectively corresponding to the case where the input optical power is equally distributed among all Xbar columns. To this end, the total Xbar power insertion losses can be obtained for every column when excluding the losses of the input modulation stage by considering $a = 1$ and can be calculated as:

$$IL_{Xbar,dB} = IL_{node,dB} + 2\log_2(N_f)\, IL_{coup,dB} + IL_{\xi,dB} + \left(\frac{N_f}{2} - 1\right) IL_{x,dB} - 10\log_{10}(\xi_1^2) + 20\log_{10}\left(\frac{N_f}{N}\right) \tag{8}$$

where $IL_{\xi,dB} = -10\log_{10}(l_\xi^2)$ and $IL_{x,dB} = -10\log_{10}(l_x^2)$ stand for the optical power loss per $\xi^2{:}\,t^2$ directional coupler and waveguide crossing, respectively. It should be underlined that $\xi_1$ can be determined based on the recursive procedure and depends only on $N_f$, $IL_{\xi,dB}$ and $IL_{x,dB}$, as described in Supplementary material, Section B. The base case and the formula for the recursive procedure are given by:

$$\xi_{M-1}^2 = \left[1 + l_x^{2\left(\frac{N_f}{2}-1-\max\{1,\log_2(N_f)-2\}\right)}\right]^{-1} \tag{9a}$$

$$\xi_c^2 = \left[1 + \left(\xi_{c+1} l_\xi l_x^{\max\{1,\log_2(N_f)-2\}}\right)^{-2}\right]^{-1}, \quad c \in [1, M-2] \tag{9b}$$

The relationship given by Eq. (8) shows that the insertion losses of the Xbar depend only linearly on the insertion losses of its node technology, $IL_{node,dB}$, whereas the exponential scaling (proportional to the Xbar size, $N_f$) takes place only with respect to a) the Y-junction splitters and combiner losses $IL_{coup,dB}$ employed at the front-end splitting and BCFS stages, and b) the waveguide crossing insertion losses $IL_{x,dB}$. Both of them, however, comprise well-established passive circuits with state-of-the-art insertion losses being as low as 0.06 dB [21] and 0.02 dB [22] for Y-junction MMI couplers and waveguide crossings, respectively. Finally, the term $20\log_{10}(N_f/N)$ reveals the additional losses encountered when the matrix dimension N is not a power of two, where an Xbar layout with $N_f = 2^{\lfloor log_2(N)\rfloor+1}$ is utilized with its $N_f - N$ rows being deactivated. Given that $N_f/N$ will never exceed the value of 2, the additional losses will be always lower than 6 dB, with the lowest-loss case being 0 dB for $N = N_f$.

The linear dependence on the MZI node insertion losses can form a significant advantage compared to the SVD-Clements layout when assuming an implementation roadmap along state-of-the-art silicon photonic fabrication capabilities, both with respect to the flexibility in the adoption of MZI node technology as well as with respect to the overall insertion losses. Retaining the exponential scaling only for, typically low-loss, passive circuitry like couplers and waveguide crossings, instead of the entire MZI node, with the loss scaling with respect to the MZI node technology being transformed into a linear relationship, can allow for the use of alternative intra-node active PS technology, not necessarily required to offer ultra-low loss performance. MZI node technology may now be selected using energy consumption or operational speed as the main criterion allowing for important energy savings or speed benefits, like, for example, in the case of the BaTiO3 electro-optic [16], PCM [14] or SiSCAP [23] PS technologies, without leading to prohibitive total insertion losses for the Xbar, as would be the case in the SVD-Clements design. A quantitative perspective of this potential can be gained when assuming the use of current silicon photonic fabrication metrics in a Xbar insertion loss analysis carried out for different MZI node technology loss values and comparing to a respective SVD-Clements design.

This is illustrated in Fig. 3 where a total loss comparison between Xbar architecture and the best- and worst-case paths of the SVD-Clements design is provided on an equal footing (i.e., with the Xbar number of outputs equalling its number of inputs, $M = N$), when assuming the use of state-of-the-art silicon photonic fabrication capabilities. This analysis considers existing silicon photonic technology with losses of $IL_{coup,dB} = 0.06$ dB for the MMI couplers used in the 3dB Y-junction splitter and combiner stages [21], $IL_{\xi,dB} = 0.1$ dB for the $\xi^2 : t^2$ optical directional couplers [24] and $IL_{x,dB} = 0.02$ dB for the waveguide crossings [22]. More specifically, Fig. 3(a) depicts the total insertion losses provided by Eqs. (1) and (2) for the best- and worst-case SVD-Clements paths (dash and dot lines, respectively) and by Eq. (8) for the Xbar design (dash-dot lines) for MZI-node loss values $IL_{node,dB}$ ranging between 0-2 dB in the case of a 4×4 (black lines) and an 8×8 (red short lines) matrix implementation. It can be clearly observed that the slope of the Xbar insertion loss is significantly lower compared to the respective slopes of the SVD-Clements design, as a direct result of the linear, as opposed to the exponential, dependence of the insertion loss on the MZI node losses. The linear dependence also leads to a constant insertion loss slope of the Xbar, with the loss itself being only increased by a constant ~3.5 dB offset as the matrix dimension scales from 4×4 to 8×8. Moreover, it results to a more loss-tolerant configuration, since the SVD-Clements layout retains a slightly lower total insertion loss only for ultra-low node losses of $IL_{node,dB} < 0.15$ dB but leads to higher losses when $IL_{node,dB}$ exceeds 0.15 dB, with the loss performance gap between the two architectures constantly increasing with $IL_{node,dB}$. The Xbar losses remain as low as 8.5 dB and 12 dB in the case of a 4×4 and an 8×8 matrix design even when $IL_{node,dB}$ equals 2 dB, while the corresponding loss values for the SVD-Clements design extend between 16-24 dB and 27-43 dB, respectively.

The scalability of the proposed Xbar layout in terms of the circuit size and its comparison with the respective SVD-Clements architecture are validated via Fig. 3(b) and (c), which illustrate the total losses with respect to the matrix dimension $N$, varying between 4 and 64, for two different values of losses per node $IL_{node,dB}$, 1 dB in Fig. 3(b) and 2 dB in Fig. 3(c). These node-loss values have been selected to comply with various state-of-the-art silicon-based PS technologies [14-16], so as to highlight the potential of the Xbar to operate with powerful non-thermo-optic MZI node technology. Both Fig. 3(b) and (c) show that the losses in dB of the SVD-Clements layout (black dash and dot lines for the best- and worst-loss path cases, respectively) increase with $N$ linearly, suggesting an exponential dependence when losses are expressed in linear scale. The slope gets higher in the case of $IL_{node,dB} = 2$ dB ,as expected from Eqs. (1) and (2). At the same time, the Xbar insertion losses (red straight lines) reveal a ramp-like behaviour owing to the loss penalty for $N < N_f$ originating from the factor $20\log_{10}(N_f/N)$ of Eq. (8), experiencing a rapid jump from 0 dB to approximately 6 dB whenever $N$ increases beyond $2^{n}$, with the loss-penalty slowly decaying until reaching 0 dB when $N = 2^{n+1}$. The Xbar insertion losses increase with $N$ at a slower rate compared to the SVD-Clements design, since they are primarily determined by the Y-junction coupler and the waveguide crossing losses and not by the MZI node losses as can be confirmed by Eq. (8). Moreover, the Xbar insertion losses are almost always lower than the respective SVD-Clements layout losses: Fig. 3(b) shows that the Xbar insertion losses extend slightly beyond the SVD-Clements best-case path losses only in the case of N=5, while remaining lower for all other values of *N,* with the loss difference between Xbar and SVD-Clements best-case increasing with $N$ and reaching 28.2 dB for N=32 and 55.6 dB for N=64. In the case of higher node losses shown in Fig. 3(c), the insertion loss advantages of the Xbar design become even higher since it offers constantly lower losses for all values of $N$ and the loss difference gets even more pronounced reaching 60.2 dB for N=32 and 119.6 dB for N=64. It should be, also, noted that the overall insertion losses remain always within a rather feasible power budget of ~30 dB even for N=64, while the respective SVD-Clements performance scales up to non-realistic power budget expectations of ~150 dB.

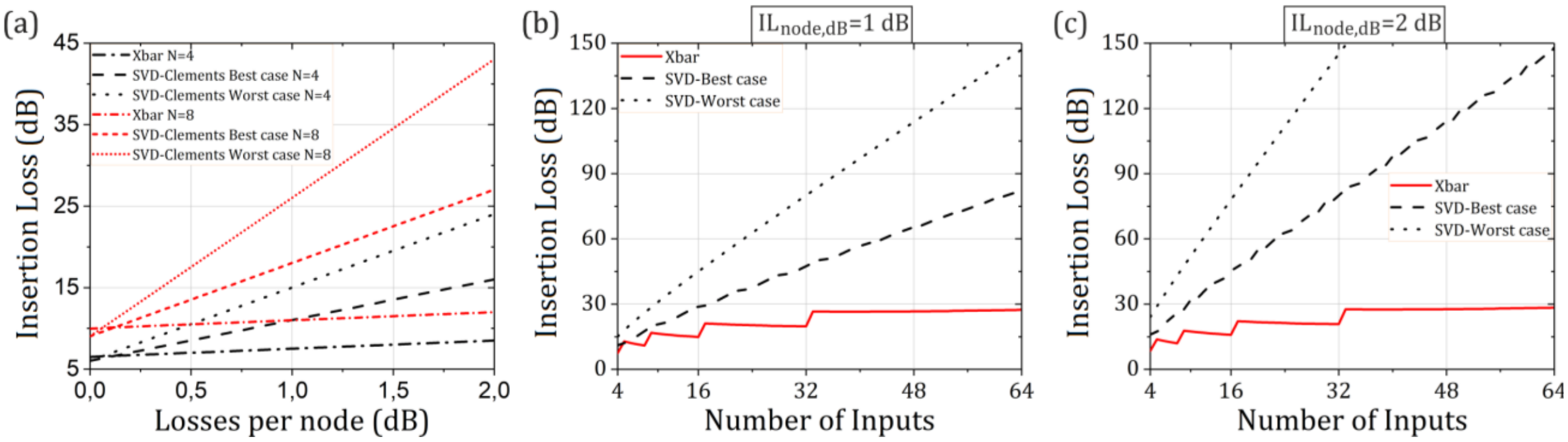

**Fig. 3.** Total insertion loss comparison between the N×N Xbar and SVD-Clements architecture (a) vs losses per node $IL_{node,dB} \in [0, 2]$ dB for N=4 (black lines, dash-dot line for the Xbar, dash for the SVD-Clements best-case and dot for the respective worst-case) and N=8 (red lines, short dash-dot line for the Xbar, short dash for the SVD-Clements best-case and short dot for the respective worst-case) and (b)-(c) vs the number of inputs $N \in [4, 64]$ when (b) $IL_{node,dB} = 1$ dB or (c) $IL_{node,dB} = 2$ dB, of the Xbar (red straight line) and the SVD-Clements (black dash line and black dot line for the best- and worst-cases, respectively) schemes.

## 5. FIDELITY ANALYSIS: LOSS- AND PHASE-ERROR TOLERANCE

Fidelity forms a highly critical factor for linear optics when lossy and/or imperfect elements are used, providing a measure of accuracy of the optical device when representing experimentally an algebraic matrix. In order to compare our Xbar design with the state-of-the-art architectures in terms of their fidelity performance, we calculate and compare the fidelity of our design with the fidelity accomplished by the SVD-Clements design for the cases of employing a) lossy optical elements, and b) phase errors in the MZI node phase shifting structures. The standard fidelity measure, based on Frobenius inner product of two matrices, is given by the following equation:

$$F(Y_{exp}, Y) = \left| \frac{tr(Y^{\dagger} Y_{exp})}{\sqrt{tr(Y^{\dagger} Y)\ tr(Y_{exp}^{\dagger} Y_{exp})}} \right|^2 \tag{10}$$

where $Y$ and $Y_{exp}$ are the N×N target matrix and its experimental implementation, respectively, with $Y^{\dagger}$ and $Y_{exp}^{\dagger}$ denoting their conjugate transposes, respectively. This measurement indicates the accuracy/error of a device when it is implemented experimentally (incorporating loss). The matrix values are enforced over electrical fields, meaning that finally the resulting electrical field considered to form the trace of the matrix product has to account also for the phase (imaginary part) of the constituent electrical fields prior applying the absolute value. The normalization employed in this fidelity definition allows us to focus, solely, on unbalanced instead of balanced losses, so that we do not distinguish between matrices that differ by just a constant multiplicative factor.

The loss-induced fidelity analysis can be carried out via analytical expressions using the Xbar matrix formulation given by Eq. (6). We show that the fidelity is unity when designing a loss-balanced configuration where the factors $\xi_c$ of the directional couplers have been selected to ensure the same losses among all columns, i.e., $p_c = p_1$ for $c \in [2, \mathrm{M}]$, transforming the transmittivity matrix into $P = p_1 I_M$, with $I_M$ standing for identity matrix of size M. With the target matrix being $Y = W^T$, the experimentally realized matrix through our Xbar architecture will equal to $Y_{exp} = P^T W^T = p_1 W^T$ according to Eq. (6). Their conjugate transposes read $Y^{\dagger} = W^*$and $Y_{exp}^{\dagger} = p_1 W^*$, with asterisk denoting conjugate matrix and assuming $p_1 = p_1^*$, since, according to Eq. (7) the coefficient $p_1$ is real. The matrix products become $Y^{\dagger} Y_{exp} = p_1 W^* W^T$, $Y^{\dagger} Y = W^* W^T$ and $Y_{exp}^{\dagger} Y_{exp} = p_1^2 W^* W^T$, and they differ only by a constant multiplicative factor given as a degree of $p_1$. Given that we are interested in the trace of the matrix product $W^* W^T$, we determine the elements along its main diagonal, which read $(W^* W^T)_{k,k} = \sum_{i=1}^{N} |w_{ki}|^2$, implying they are real, yielding also a real trace $\mathrm{tr}(W^* W^T) = \sum_{k=1}^{N} \sum_{i=1}^{N} |w_{ki}|^2$, as can be also devised based on Frobenius inner product. Having all quantities of interest being real, we conclude to:

$$F_{Xbar} = F(Y_{exp}, Y) = \left| \frac{p_1 \mathrm{tr}(W^* W^T)}{\sqrt{\mathrm{tr}(W^* W^T) p_1^2\ \mathrm{tr}(W^* W^T)}} \right|^2 = 1 \tag{11}$$

The design of a loss-balanced layout, however, requires high accuracy and high resolution in the values of the $\xi_c$, which may not always be possible, especially for high number of Xbar columns, given the current technology and fabrication limitations. Given that $P$ is a diagonal matrix with its diagonal elements in practice differing among themselves and from the unity value, the fidelity obtained through Eq. (10) will generally be lower than unity when an unbalanced loss design is employed. However, the Xbar layout allows - even in this case - for a completely restorable fidelity performance in direct contrast with the SVD-Clements design where fidelity restoration cannot be accomplished. The Xbar fidelity restoration can be achieved by defining a new matrix $Y'_{exp} = (P^T)^{-1} Y_{exp} = (P^T)^{-1} P^T W^T = W^T = Y$ where fidelity will always equal unity, i.e., $F(Y'_{exp}, Y) = 1$, requiring, in practice, a simple procedure of multiplying the Xbar's output by the inverse of its $P^T$ diagonal matrix. Multiplication with this new diagonal matrix from the left side effectively corresponds to the introduction of differentiated factor at every column output, i.e., by inserting a factor of $1/p_c$ at every column output $c$. This yields a loss-balanced Xbar architecture where the power balancing is achieved by pondering the outputs and thus compensating for the power distribution inequality due to the non-optimal splitting ratios. Even if all $\xi_c^2 : t_c^2$ optical couplers across columns are assumed to have the same coupling ratios $\xi_c$ and $t_c$, which would reduce design and fabrication complexity, the lack of an a-priori loss-balanced scheme is easily compensated at the Xbar's output.

Assuming, for example, that $\xi_c = \xi_1$ and $t_c = t_1$ for every $c \in [1, \mathrm{N}-1]$ in the case of an N×N matrix arrangement, then $p_c/p_{c-1} = L_c t_{c-1}\xi_c / L_{c-1}\xi_{c-1} = t_1 l_\xi l_x^{\max\{1,\log_2(N_f)-2\}}$, suggesting that the differential loss between two successive columns is equal to the losses of $\max\{1, \log_2(N_f)-2\}$ waveguide crossings employed at every row of the BCFS stage multiplied by the excess losses of a directional coupler $\xi^2 : t^2$ and the coupling coefficient $t_1$. The exception is the last column N, where the non-existent coupler is described by $\xi_N = 1$ and we have $p_N/p_{N-1} = L_N t_{N-1}\xi_N / L_{N-1}\xi_{N-1} = \left(t_1/\sqrt{1-t_1^2}\right) l_x^{\max\{1,\log_2(N_f)-2\}-\left(\frac{N_f}{2}-1\right)}$.
Finally, the fidelity restoration matrix $(P^T)^{-1}$, in this case, can be defined by its diagonal elements, where the $c$-th element $p_c^{-1}$ can be calculated as:

$$p_c^{-1} = p_1^{-1} \cdot \begin{cases} t_1^{-(c-1)} l_\xi^{-(c-1)} l_x^{-(c-1)\max\{1,\log_2(N_f)-2\}}, & c \in [1, N-1] \\ \sqrt{1-t_1^2}\, t_1^{-(N-1)} l_\xi^{-(N-2)} l_x^{-(N-1)\max\{1,\log_2(N_f)-2\}+\left(\frac{N_f}{2}-1\right)}, & c = N \end{cases} \quad (12)$$

This proves that our Xbar architecture supports a simple fidelity restoration mechanism requiring just the introduction of simple attenuation or gain elements at its column outputs.

This fidelity restoration mechanism is a significant advantage compared to SVD-based matrix designs, as it allows to accurately implement any target matrix through simple loss-balancing at the device output. On the contrary, SVD-based architectures can turn into loss-balanced configurations only when intervening in their inner-design and equipping every of their cascaded MZI stages with dummy MZIs [25-26], which effectively means that all possible optical paths turn into equal-loss paths that experience the same loss with what has been considered as the worst-case optical loss path in our previous analysis. To this end, loss-balancing in the SVD-based designs can only be accomplished during device design and at the expense of much higher insertion losses, significantly increasing the gap between insertion losses of the SVD compared to the Xbar configuration, as has been already analyzed through Fig. 3.

The quantified benefits provided by the loss-induced fidelity restoration mechanism of the Xbar over the SVD-Clements layout are revealed in Fig. 4. Having proven that the Xbar retains fidelity of 1 regardless of node losses, we perform fidelity calculations for the SVD-Clements design using Monte-Carlo method with 500 random target matrices $Y$ generated for every given number of inputs $N$, further decomposed to sets of $(\mathrm{U}, \Sigma, \mathrm{V}^\dagger)$ (more details in Section C of the Supplementary material). Fig. 4(a) depicts the fidelity performance for two specific matrix dimensions of 4×4 and 8×8, when the losses of every MZI node vary between 0-2 dB. Fidelity degradation in the SVD-Clements layout becomes higher as the MZI node losses increase and the effect becomes even more pronounced when scaling the matrix dimension from 4×4 towards 8×8. This suggests that high fidelity performance in SVD-Clements-based schemes can, only, be obtained when ultra-low loss PS technologies with loss values <0.1 dB are used, as in the case of thermo-optic (T/O) MZIs [27] or Ring Resonators (RRs) [28], where, however, a power consumption of several mWs is required and only MHz-scale operational speeds can be sustained. Higher-loss phase shifting technologies would lead to significant fidelity degradation, implying that a whole range of highly promising technologies should be probably excluded from their employment in SVD-based schemes, such as state-of-the-art PCM and silicon-based electro-optic PS technology [14-16], where losses can exceed 1 dB per node.

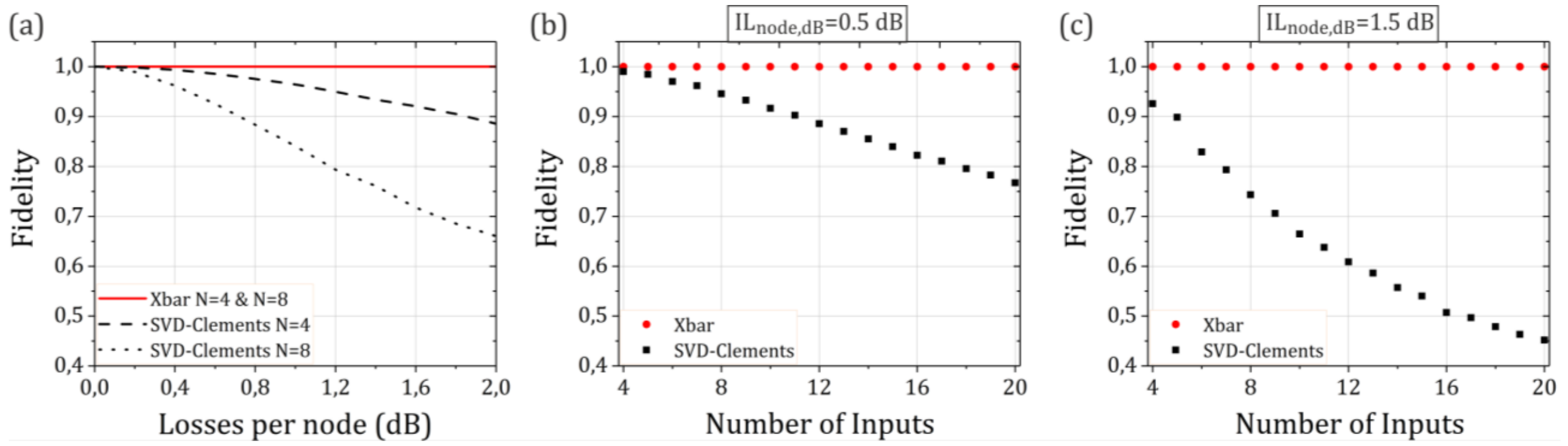

**Fig. 4**. Fidelity comparison of Xbar vs SVD-Clements schemes with respect to (a) the losses per node $IL_{node,dB} \in [0, 2]$ dB for a number of inputs N=4 (solid line for Xbar and dashed line for SVD-Clements) and N=8 (solid line for Xbar and dot line for SVD-Clements) and (b)-(c) the number of inputs $N \in [4, 20]$ for a constant loss of (b) $IL_{node,dB} = 0.5$ dB and (c) $IL_{node,dB} = 1.5$ dB per node.

The fidelity performance gap between the SVD-Clements design and our architecture becomes dramatically wider when higher dimension layouts are implemented. This can be, also, confirmed by Fig. 4(b) and (c), where the fidelity performance with respect to the circuit size for two different MZI node loss values of 0.5 dB (Fig. 4(b)) and 1.5 dB (Fig. 4(c)) is illustrated. The SVD-Clements fidelity falls below 80% even for matrix dimensions as low as 18×18 and 7×7 when 0.5 dB and 1.5 dB losses per MZI node are employed, with the 20×20 matrix layouts yielding fidelity performance ~76% and ~45% for the two respective node loss cases.

In addition to the loss-induced fidelity degradation, $\theta$ and $\varphi$ phase errors encountered at the Xbar node PSs can also critically affect the performance and lead to accuracy deviations in the experimental realization of a target matrix. This is also expected to be the case for the SVD-Clements architecture, given that phase-error-induced fidelity degradation has been already analyzed for unitary matrices relying on the Clements and Reck architectures [29]. In order to evaluate the phase-error tolerance of our Xbar architecture in comparison with the SVD-Clements layout, we employ again the standard fidelity measure of Eq. (10) and, using the Monte Carlo method, we calculate the respective fidelity metrics for the Xbar and the SVD-Clements designs by assuming lossless architectures for 500 arbitrary target matrices generated for every given matrix dimension N (more details in Section C of the Supplementary material).

Figure 5 illustrates the fidelity comparison of the two architectures. Figure 5(a) and (b) depict the fidelity performance for a 4×4 and an 8×8 matrix implementation, respectively, when the phase-error standard deviation $\sigma$ ranges between 0-0.2 rad. It can be observed that the Xbar architecture outperforms the SVD-Clements design in both cases, yielding a slightly reduced performance when increasing the matrix dimensions from 4×4 to 8×8, but retaining always a fidelity value that is greater than 97%, even in the extreme case where the standard deviation equals 0.2 rad. On the other hand, the fidelity of the SVD-Clements scheme degrades faster for higher standard deviation values, with the degradation becoming even more distinct when moving from the 4×4 to the 8×8 matrix implementation, yielding a fidelity value <87% when the standard deviation equals 0.2 rad. The effect of matrix dimensions on the fidelity performance can be more clearly witnessed in Fig. 5(c), where fidelity for both schemes has been plotted for a constant standard deviation value of 0.1 rad and for matrix dimension $N$ ranging between 4 and 20. The Xbar architecture reveals an almost constant fidelity value as $N$ increases, allowing fidelity to remain above 99% even when N=20, whereas the respective SVD-Clements design shows a significant fidelity drop down to only 87%. This verifies that the Xbar architecture shows a stronger phase-error-resistive behaviour compared to the SVD-Clements scheme, with its fidelity experiencing a weaker dependence both on phase-error standard deviation and on matrix dimension scaling. This can be explained by the one-to-one mapping between the target matrix elements and Xbar nodes in our configuration, whereas, in the case of the SVD-Clements design, the mapping is not bijective, requiring a product of transfer matrices from multiple MZI nodes to achieve target matrix element. As such, assuming that a phase-error is experienced at a single (θ,φ) phase pair within a node, this will affect a single node and a single matrix element in the Xbar layout, as opposed to the multiple matrix entries, as will be the case in the SVD-Clements scheme. Moreover, given that the number of matrix entries affected by a phase error at a single (θ,φ) phase pair increases with matrix dimension as a result of the higher number of cascaded MZI nodes being employed, the impact of error becomes much stronger in the SVD-Clements scheme as the circuit size increases, as it can be seen also in Fig. 5(c). On the contrary, the independent nature of Xbar's nodes and their one-to-one correspondence to target matrix elements is retained irrespective of the matrix size, obviously resulting to a more phase-error tolerant design for increasing circuit sizes.

Table 1 summarizes the analysis reported in the above sections and provides an overview of the performance characteristics of the Xbar and the SVD-Clements architectures.

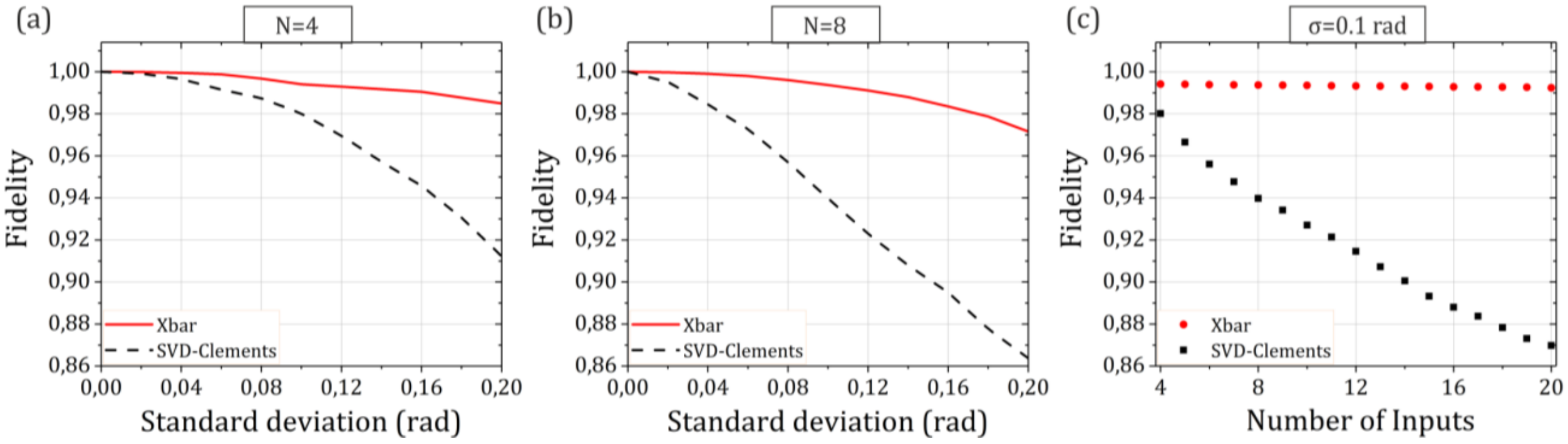

**Fig. 5**. Fidelity comparison of Xbar (red straight-lines/dots) vs SVD-Clements (black dashed-lines/squares) design with respect to (a)-(b) the standard deviation $\sigma \in [0, 0.2]$ rad for a number of inputs (a) N=4, (b) N=8 and (c) the number of inputs $N \in [4, 20]$ for a constant standard deviation of σ = 0.1 rad.

*Table 1 Xbar and SVD-Clements architectures performance summary*

| Performance parameter | Xbar | SVD-Clements |
|---|---|---|
| Insertion loss dependence on loss per node | Linear | Exponential |
| Loss-induced fidelity | Fully Restorable | Significant degradation |
| Phase-error-induced fidelity | Small degradation | Significant degradation |
| Scalability | High | Low |
| Versatility to node technology | High | Low |
| Programming complexity (for an N×N matrix) | O(1) | O(N) |

## 5. CONCLUSION

We have demonstrated a novel coherent Xbar architecture that can serve as a universal linear operator, being capable of implementing any real- and/or complex-valued matrix. This is, to the best of our knowledge, the first linear operator that can be experimentally realized without relying on state-of-the-art SVD-based techniques, building upon a layout that allows to directly represent a single matrix element via the amplitude and phase values of a single Xbar node. This Xbar architecture avoids the use of cascaded MZI nodes as being typically required in SVD-based designs, allowing in this way for significantly lower insertion losses and improved scalability compared to state-of-the-art implementations. A detailed comparative analysis with respect to the so far optimal SVD-Clements layout has been performed, revealing that our circuit allows for an a priori loss-balanced configuration through appropriate design of the constituent components and, most importantly, is the first universal linear operator that supports a simple fidelity restoration mechanism even when the losses have not been balanced among the different Xbar columns. Moreover, it outperforms the SVD-based schemes even in the presence of phase-errors, allowing for a significantly more loss- and phase-error tolerant configuration. All these advantages can turn into significant functional improvements, since it can support also alternative and higher-loss MZI node technologies that could equip linear optical operators with significant speed and energy consumption benefits compared to the SVD-based schemes. To the best of our knowledge, this work introduces for the first time a multi-port interferometric design that can serve as an any matrix operator without adopting SVD factorization and unitary linear optics, concluding to a practical, lower-loss and highly robust architecture that can be utilized as a universal linear optical circuit in a diverse range of application fields.

## References

1. David A. B. Miller, "Self-configuring universal linear optical component [Invited]," Photon. Res. 1, 1-15 (2013).
2. J. Carolan, C. Harrold, C. Sparrow, E. Martín-López, N. J. Russell, J. W. Silverstone, P. J. Shadbolt, N. Matsuda, M. Oguma, M. Itoh, G. D. Marshall, M. G. Thompson, J. C. F. Matthews, T. Hashimoto, J. L. O'Brien, and A. Laing, "Universal linear optics," Science 349, 711–716 (2015).
3. Y. Shen, N. C. Harris, S. Skirlo, M. Prabhu, T. Baehr-Jones, M. Hochberg, X. Sun, S. Zhao, H. Larochelle, D. Englund, and M. Soljačić, "Deep learning with coherent nanophotonic circuits," Nat. Photonics 11, 441–446 (2017).
4. A. N. Tait, M. A. Nahmias, B. J. Shastri and P. R. Prucnal, "Broadcast and Weight: An Integrated Network For Scalable Photonic Spike Processing," in Journal of Lightwave Technology, vol. 32, no. 21, pp. 4029-4041, 1 Nov.1, 2014, doi: 10.1109/JLT.2014.2345652.
5. Capmany, J., Gasulla, I. & Pérez, D. The programmable processor. Nature Photon 10, 6–8 (2016).
6. Lin Yang, Ruiqiang Ji, Lei Zhang, Jianfeng Ding, and Qianfan Xu, "On-chip CMOS-compatible optical signal processor," Opt. Express 20, 13560-13565 (2012).
7. H. Zhang, M. Gu, X.D. Jiang, J. Thompson, H, Cai, S. Paesani, A. Laing, Y. Zhang, M. H. Yung, Y. Z. Shi, F. K. Muhammad, G. Q. Lo, X. S. Luo, B. Dong, D. L. Kwong, L. C. Kwek, and A. Q. Liu," An optical neural chip for implementing complex-valued neural network", Nat. Commun. 12, 457 (2021). https://doi.org/10.1038/s41467-020-20719-7
8. L. Zhuang, C. G. H. Roeloffzen, M. Hoekman, K.-J. Boller, and A. J. Lowery, "Programmable photonic signal processor chip for radiofrequency applications," Optica 2, 854–859 (2015).
9. D. Pérez, I. Gasulla, J. Capmany, and R. A. Soref, "Reconfigurable lattice mesh designs for programmable photonic processors," Opt. Express 24, 12093–12106 (2016).
10. M. Miscuglio, V.J. Sorger, "Photonic Tensor cores for machine learning", Appl. Phys. Rev., vol.7 issue 3 (2020).
11. Nicholas C. Harris, Gregory R. Steinbrecher, Mihika Prabhu, Yoav Lahini, Jacob Mower, Darius Bunandar, Changchen Chen, Franco N. C. Wong, Tom Baehr-Jones, Michael Hochberg, Seth Lloyd & Dirk Englund, "Quantum transport simulations in a programmable nanophotonic processor", Nature Photonics 11, 447-452 (2017)
12. M. Reck, A. Zeilinger, H. J. Bernstein, and P. Bertani, "Experimental realization of any discrete unitary operator," Phys. Rev. Lett. 73, 58–61 (1994).
13. W. R. Clements, P. C. Humphreys, B. J. Metcalf, W. Steven Kolthammer and I. A. Walmsley, "Optimal design for universal multiport interferometers," Optica 3, 1460–1465 (2016).

14. W. Jiang, “Nonvolatile and ultra-low-loss reconfigurable mode (De)multiplexer/switch using triple-waveguide coupler with Ge2Sb2Se4Te1 phase change material,” Sci. Rep., vol. 8, no. 1, pp. 1–12, 2018.
15. A. Manolis, J. Faneca, T. D. Bucio, A. Baldycheva, A. Miliou, F. Y. Gardes, N. Pleros, and C. Vagionas, "Non-volatile integrated photonic memory using GST phase change material on a fully etched Si3N4/SiO2 waveguide," in Conference on Lasers and Electro-Optics, OSA Technical Digest (Optical Society of America, 2020), paper STh3R.4
16. Felix Eltes, Christian Mai, Daniele Caimi, Marcel Kroh, Youri Popoff, Georg Winzer, Despoina Petousi, Stefan Lischke, J. Elliott Ortmann, Lukas Czornomaz, Lars Zimmermann, Jean Fompeyrine, and Stefan Abel, "A BaTiO3-Based Electro-Optic Pockels Modulator Monolithically Integrated on an Advanced Silicon Photonics Platform," J. Lightwave Technol. 37, 1456-1462 (2019).
17. G. Mourgias-Alexandris, A. Totovic, A. Tsakyridis, N. Passalis, K. Vyrsokinos, A. Tefas, N. Pleros, "Neuromorphic Photonics With Coherent Linear Neurons Using Dual-IQ Modulation Cells," in Journal of Lightwave Technology, vol. 38, no. 4, pp. 811-819, 15 Feb.15, (2020).
18. S. Pai, I. A. D. Williamson, T. W. Hughes, M. Minkov, O. Solgaard, S. Fan, D. A. B. Miller, "Parallel Programming of an Arbitrary Feedforward Photonic Network," in IEEE Journal of Selected Topics in Quantum Electronics, vol. 26, no. 5, pp. 1-13, Sept.-Oct. 2020, Art no. 6100813
19. J. Feldmann, M. Youngblood, M. Karpov, H. Gehring, X. Li, M. Stappers, M. Le Gallo, X. Fu, A. Lukashchuk, A.S. Raja, J. Liu, C.D. Wright, A. Sebastian, T.J. Kippenberg, W.H.P. Perince and H. Bhaskaran, “Parallel convolutional processing using an integrated photonic tensor core”. Nature 589, 52–58 (2021).
20. S. Ohno, K. Toprasertpong, S. Takagi and M. Takenaka, “Demonstration of Classification Task Using Optical Neural Network Based on Si microring Resonator Crossbar Array”, ECOC proceedings, (2020).
21. Z. Sheng et al., "A Compact and Low-Loss MMI Coupler Fabricated With CMOS Technology," in IEEE Photonics Journal, vol. 4, no. 6, pp. 2272-2277, Dec. 2012, doi: 10.1109/JPHOT.2012.2230320.
22. Yangjin Ma, Yi Zhang, Shuyu Yang, Ari Novack, Ran Ding, Andy Eu-Jin Lim, Guo-Qiang Lo, Tom Baehr-Jones, and Michael Hochberg, "Ultralow loss single layer submicron silicon waveguide crossing for SOI optical interconnect," Opt. Express 21, 29374-29382 (2013).
23. M. Webster et al., "An efficient MOS-capacitor based silicon modulator and CMOS drivers for optical transmitters," 11th International Conference on Group IV Photonics (GFP), Paris, 2014, pp. 1-2, doi: 10.1109/Group4.2014.6961998.
24. B. Sharma, K. Kishor, A. Pal, S. Sharma and R. Makkar, “Design and simulation of ultra-low loss triple tapered asymmetric directional coupler at 1330nm”, in Elsevier Microelectronics Journal, vol. 107 (2021).
25. Farhad Shokraneh, Simon Geoffroy-gagnon, and Odile Liboiron-Ladouceur, "The diamond mesh, a phase-error- and loss-tolerant field-programmable MZI-based optical processor for optical neural networks," Opt. Express 28, 23495-23508 (2020).
26. D. A. B. Miller, “Perfect optics with imperfect components,” Optica 2, 747–750 (2015).
27. N. C. Harris, Y. Ma, J. Mower, T. Baehr-Jones, D. Englund, M. Hochberg, and C. Galland, “Efficient, compact and low loss thermo-optic phase shifter in silicon,” Opt. Express, vol. 22, no. 9, pp. 10 487–10 493, May 2014. [Online]. Available: http: www.opticsexpress.org/abstract.cfm?URI=oe-22-9-10487
28. C. Sun, M. Wade, M. Georgas, S. Lin, L. Alloatti, B. Moss, R. Kumar, A. H. Atabaki, F. Pavanello, J. M. Shainline, J. S. Orcutt, R. J. Ram, M. Popovic, and V. Stojanovic, “A 45 nm cmos-soi monolithic photonics platform with bit-statistics-based resonant microring thermal tuning,” IEEE Journal of Solid-State Circuits, vol. 51, no. 4, pp. 893–907, 2016.
29. Roel Burgwal, William R. Clements, Devin H. Smith, James C. Gates, W. Steven Kolthammer, Jelmer J. Renema, and Ian A. Walmsley, "Using an imperfect photonic network to implement random unitaries," Opt. Express 25, 28236-28245 (2017).

## SUPPLEMENTARY MATERIAL

### *A. MATHEMATICAL ANALYSIS OF THE CROSSBAR’S OPERATION*

In order to calculate the electric field at the output of each $N_f$×M Xbar’s column, let us first define $a \le 1$ as the electric field transmission coefficient of the input amplitude modulator $x_r$ when operated in transparency. Additionally, let us define $l_{coup} \le 1$ and $k \le 1$ as the electric field transmission coefficients of a 3dB coupler and a Phase Shifter (PS), respectively, where $IL_{coup,dB} = -10\log_{10}\left(l_{coup}^2\right)$ and $IL_{PS,dB} = -10\log_{10}(\mathrm{k}^2)$ denote the corresponding optical power insertion losses. In each MZI-node, where two 3dB couplers and two PSs θ, φ are employed, the total electric field transmittivity factor is $T_{node} = l_{coup}^2 k^2 \le 1$. Moreover, we define, $l_\xi \le 1$ as the electric field transmission coefficient of the $\xi^2$:$t^2$ coupler and $l_x \le 1$ as the electric field transmission coefficient associated with the losses per crossing, so that $IL_{\xi,dB} = -10\log_{10}\left(l_\xi^2\right)$ and $IL_{x,dB} = -10\log_{10}(l_x^2)$. After balancing the losses coming from different number of waveguide crossings within an Xbar column by adding dummy waveguide crossings in the BCFS circuit sections to amount to $\max\{1, \log_2 N_f - 2\}$ along the forwarding waveguides and $N_f/2 - 1$ along each path of the vertical coupling section for $N_f \ge 2^2$, or, equivalently, $N > 2$, as described in the main body of the manuscript, we define the total passive circuit transmission factor of the $c$-th column $L_c \le 1$. Assuming $N > 2$, the total passive loss per column can be calculated as follows:

$$L_1 = l_{coup}^{2\,log_2(N_f)} l_\xi l_x^{\frac{N_f}{2}-1}$$

$$L_2 = l_{coup}^{2\,log_2(N_f)} l_\xi^2 l_x^{\frac{N_f}{2}-1+max\{1,log_2(N_f)-2\}}$$

$$L_3 = l_{coup}^{2\,log_2(N_f)} l_\xi^3 l_x^{\frac{N_f}{2}-1+2\,max\{1,log_2(N_f)-2\}}$$

$$\vdots$$

$$L_{M-1} = l_{coup}^{2\,log_2(N_f)} l_\xi^{M-1} l_x^{\frac{N_f}{2}-1+(M-2)\,max\{1,log_2(N_f)-2\}}$$

$$L_M = l_{coup}^{2\,log_2(N_f)} l_\xi^{M-1} l_x^{(M-1)\,max\{1,log_2(N_f)-2\}}$$

or:

$$L_c = \begin{cases} l_{coup}^{2\,log_2(N_f)}\, l_\xi^c\, l_x^{\frac{N_f}{2}-1+(c-1)\,max\{1,log_2(N_f)-2\}}, & c \in [1, M-1], N > 2 \\ l_{coup}^{2\,log_2(N_f)}\, l_\xi^{M-1}\, l_x^{(M-1)\,max\{1,log_2(N_f)-2\}}, & c = M, N > 2 \end{cases} \tag{S1a}$$

where $N_f = 2^n$, for $n = \lfloor log_2(N) \rfloor + 1$, as analyzed in the main manuscript. In the special case of $N = 2$, no waveguide crossings exist, implying:

$$L_c = \begin{cases} l_{coup}^{2\,log_2(N_f)} l_\xi^c, & c \in [1, M-1], N = 2 \\ l_{coup}^{2\,log_2(N_f)} l_\xi^{M-1}, & c = M, N = 2 \end{cases} \tag{S1b}$$

Consequently, the electric field at every column output can be obtained through the relation (S2):

$$E_{out,c} = \alpha T_{node} L_c \frac{1}{N_f} \left( \prod_{q=1}^{c-1} t_q \right) \xi_c \left( \sum_{r=1}^{N} x_r w_{r,c} e^{j\varphi_{r,c}} \right) E_{in} \tag{S2}$$

The product $\prod_{q=1}^{c-1} t_q$ represents the total transmission coefficient of all $\xi^2 : t^2$ couplers preceding the column $c$, and, by convention, in case of the empty set is equal to 1, i.e., for the 1-st column, $\prod_{q=1}^{0} t_q = 1$. In case of the last column, the splitting ratio can be defined by $\xi_M = 1$, as no coupler exists and the signal is only forwarded to the last Xbar column by the previous, (M-1)-th column's splitting stage.

The output electric field given by (S2) can be written in matrix form as:

$$E_{out} = (WP)^T X E_{in} = P^T W^T X E_{in} \tag{S3}$$

where $E_{out} = \left[E_{out,1}, E_{out,2}, \dots, E_{out,M}\right]^T$ stands for the output vector, $E_{in}$ for the input optical signal, $X = [x_1, x_2, \dots, x_N]^T$ the vector of inputs,

$$W = \begin{bmatrix} w_{1,1}e^{j\varphi_{1,1}} & w_{1,2}e^{j\varphi_{1,2}} & \cdots & w_{1,M}e^{j\varphi_{1,M}} \\ w_{2,1}e^{j\varphi_{2,1}} & w_{2,2}e^{j\varphi_{2,2}} & \cdots & w_{2,M}e^{j\varphi_{2,M}} \\ \vdots & \vdots & \ddots & \vdots \\ w_{N,1}e^{j\varphi_{N,1}} & w_{N,2}e^{j\varphi_{N,2}} & \cdots & w_{N,M}e^{j\varphi_{N,M}} \end{bmatrix} \tag{S4}$$

the targeted real/complex-valued matrix, implemented via the MZM-nodes of the Xbar and, finally, by $P$ we denote an M×M diagonal matrix $P = \text{diag}[p_1, p_2, \dots, p_M]$, encompassing all the transmission factors that correspond to the insertion losses and splitting ratios. The elements of the $P$ matrix can be calculated by Eq. (S5):

$$p_c = \alpha T_{node} L_c \frac{1}{N_f} \left( \prod_{q=1}^{c-1} t_q \right) \xi_c \tag{S5}$$

## *B. TOTAL INSERTION LOSS CALCULATION IN THE CROSSBAR ARCHITECTURE*

The optical power can be equally distributed among all columns of the Xbar when enforcing $p_1 = p_2 = \cdots = p_M$, implying that the coupling coefficients $\xi_c$, $t_c$ have to satisfy the following relations (S6):

$$L_c\left(\prod_{q=1}^{c-1} t_q\right)\xi_c = const. \tag{S6}$$

When $N > 2$, previous relation can be expanded to

$$\begin{aligned}\xi_1 l_\xi &= t_1\xi_2 l_\xi^2 l_x^{max\{1,log_2(N_f)-2\}} = t_1 t_2 \xi_3 l_\xi^3 l_x^{2\,max\{1,log_2(N_f)-2\}} = \cdots \\ &= t_1 t_2 \dots t_{M-2}\xi_{M-1} l_\xi^{M-1} l_x^{(M-2)\,max\{1,log_2(N_f)-2\}} \\ &= t_1 t_2 \dots t_{M-1} l_\xi^{M-1} l_x^{(M-1)\,max\{1,log_2(N_f)-2\}-\left(\frac{N_f}{2}-1\right)}\end{aligned} \tag{S7}$$

In addition to the previous set of equations, another set holds true, owing to the power conservation constrains. Namely, as the loss of the couplers has been treated independently from their splitting/coupling ratio, through the parameter $l_\xi$, the sum of the two coefficients defining the power splitting ratio must, at all times, fulfill $t_c^2 + \xi_c^2 = 1$ for $c \in [1, M-1]$. Focusing on the last two columns (M − 1 and M), that share the (M-1)-th coupler, we have:

$$\xi_{M-1} = t_{M-1} l_x^{max\{1,log_2(N_f)-2\}-\left(\frac{N_f}{2}-1\right)}, \qquad t_{M-1}^2 + \xi_{M-1}^2 = 1$$

$$\xi_{M-1}^2 = \frac{1}{1 + l_x^{2\left[\left(\frac{N_f}{2}-1\right)-max\{1,log_2(N_f)-2\}\right]}} \tag{S8}$$

Same procedure can be extended to any two consecutive columns, e.g., $c$-th and $(c+1)$-th, for $c \in [1, M-2]$ yielding:

$$\xi_c = t_c \xi_{c+1} l_\xi l_x^{max\{1,log_2(N_f)-2\}}, \qquad t_c^2 + \xi_c^2 = 1$$

$$\xi_c^2 = \frac{1}{1 + \left(\xi_{c+1} l_\xi l_x^{max\{1,log_2(N_f)-2\}}\right)^{-2}} \tag{S9}$$

Applying the recursive procedure based on the base case of $\xi_{M-1}^2$ and the recursive step defined above, all splitting coefficients can be determined. Moreover, enforcing identical coefficients $p_c$, allows us to express any column's output via the splitting coefficient $\xi_1$, which itself is a function of $l_\xi$, $l_x$ and $N_f$, as:

$$E_{out,c} = p_c\left(\sum_{r=1}^{N} x_r w_{r,c} e^{j\varphi_{r,c}}\right)E_{in} = p_1\left(\sum_{r=1}^{N} x_r w_{r,c} e^{j\varphi_{r,c}}\right)E_{in}$$

$$E_{out,c} = \alpha T_{node} L_1 \frac{1}{N_f}\xi_1\left(\sum_{r=1}^{N} x_r w_{r,c} e^{j\varphi_{r,c}}\right)E_{in} \tag{S10}$$

In order to calculate the overall loss of the Xbar, we consider lossless input modulators ($\alpha = 1$) and assume that all input modulators and all weighting nodes operate at their transparency, i.e., $x_r = w_{r,c} = 1$ and $\varphi_{r,c} = 0, \forall r, c$, which implies:

$$\alpha\left(\sum_{r=1}^{N} x_r w_{r,c} e^{j\varphi_{r,c}}\right) = N \tag{S11}$$

and, hence, we get:

$$P_{out,c} = E_{out,c}E_{out,c}^{\ *} = {p_1}^2 N^2 P_{in} \tag{S12}$$

Considering $IL_{node,dB} = -10\log_{10}(T_{node}^2)$ as the insertion loss of each MZI-node (2 couplers and 2 PSs), as in the case of SVD-Clements architecture, the total insertion loss of the Xbar can be expressed as:

$$IL_{Xbar,dB} = -10\,log_{10}\left(\frac{P_{out,c}}{P_{in}}\right) \Rightarrow$$

$$IL_{Xbar,dB} = -10\,log_{10}({p_1}^2 N^2) = -10\,log_{10}\left({T_{node}}^2 {L_1}^2 \frac{1}{N_f^2}{\xi_1}^2 N^2\right)$$

$$IL_{Xbar,dB} = -10\,log_{10}\left(T_{node}{}^{2} l_{coup}^{4log_2(N_f)} l_{\xi}^{2} l_{x}{}^{2\left(\frac{N_f}{2}-1\right)} \frac{1}{N_f^2} \xi_1{}^{2} N^2\right) \Rightarrow$$

$$IL_{Xbar,dB} = IL_{node,dB} + 2\,log_2(N_f)\,IL_{coup,dB} + IL_{\xi,dB} + \left(\frac{N_f}{2} - 1\right) IL_{x,dB} - 10\,log_{10}(\xi_1^2) + 20\,log_{10}\left(\frac{N_f}{N}\right) \quad \text{(S13)}$$

It should be underlined that $\xi_1$ can be determined based on the previously established recursive procedure and depends only on $N_f$, $IL_{\xi,dB}$ and $IL_{x,dB}$.

## *C. CROSSBAR AND SVD-CLEMENTS FIDELITY CALCULATION METHOD*

We compare the fidelity (Eq. (10) of main manuscript) of our Xbar with the fidelity accomplished by the SVD-Clements design for the cases of employing a) lossy optical elements, and b) phase errors in the MZI node phase shifting structures. Fidelity calculations have been performed using the Monte-Carlo method, where, for a given number of inputs N, we initially generate 500 random target matrices $Y$. In case of Xbar, the elements of the matrix should be mapped to the designated node modulator and phase shifter as given, whereas in the SVD-Clements case, the matrices are subsequently decomposed using SVD, producing the $U$, $\Sigma$ and $V^{\dagger}$ matrices. All elements of each matrix ($U$, $\Sigma$, $V^{\dagger}$) are then normalized by being divided by the maximum norm (magnitude) among its elements. We then calculate the corresponding θ and φ phases of each lossless MZI node for the decomposition of $U$ and $V^{\dagger}$ via the Clements method and of the MZI nodes of $\Sigma$.

In case of the lossy-elements SVD-Clements fidelity evaluation case, these phase values are enforced onto lossy MZI nodes constructing the $Y_{exp}$ matrix, required for the fidelity calculation. We use a rather simple loss model for the SVD-Clements architecture where all MZI nodes are assumed to have the same insertion losses, considering a fixed $IL_{coup,dB} = 0.06dB$ loss for each 3dB MMI coupler and $-10\log_{10}(k^2)$ loss for every PS, determined such that in total we reach the targeted value of the total node insertion losses $IL_{node,dB}$. In the case of Xbar's loss-induced fidelity, we prove via analytical expressions that it can always 1.

Regarding the phase-tolerance fidelity, for each arbitrary matrix $Y$, we generate 100 two-element sets of random phase deviation values corresponding to (θ,φ) that follow a Gaussian distribution with a mean value of zero and a standard deviation $\sigma \in [0, 0.2\,]$ rad. Fidelity is then calculated for the SVD-Clements architecture by applying the generated phase deviations onto the ideal θ and φ values, enforcing these new phase values onto the respective MZI nodes within the SVD-Clements design and concluding to the experimental matrix representation $Y_{exp}$, which is subsequently used in the fidelity calculation Eq. (10) of the main manuscript. Thereafter, following the same procedure, matrices and sets of phase deviations, we obtained the respective fidelity values for the Xbar architecture.